\documentclass[article]{elsarticle}
\usepackage[utf8]{inputenc}
\usepackage{hyperref}
\usepackage{graphicx}
\usepackage{amsmath}
\usepackage{amssymb,mathrsfs}
\usepackage[layout=margin,innerlayout=inline]{fixme}
\usepackage{tikz}
\usepackage{caption}
\usepackage{subcaption}
\usepackage{hyperref}
\usepackage{makecell}
\usepackage{multirow}
\usepackage{fancyhdr}
\makeatletter
\def\ps@pprintTitle{%
 \let\@oddhead\@empty
 \let\@evenhead\@empty
 \def\@oddfoot{\centerline{\thepage}}%
 \let\@evenfoot\@oddfoot}
\makeatother
\journal{arXiv Preprint}

\begin{document}

\begin{frontmatter}
\title{AVRA: Automatic Visual Ratings of Atrophy from MRI images using Recurrent Convolutional Neural Networks.}

%\author[]{Gustav Mårtensson\corref{mycorrespondingauthor}\fnref{myfootnote}}%
\author[ki]{Gustav Mårtensson\corref{mycorrespondingauthor}}%
\ead{gustav.martensson@ki.se}
\author[ki]{Daniel Ferreira }
\author[ki2,ks]{Lena Cavallin }
\author[ki]{J-Sebastian Muehlboeck }
\author[ki]{Lars-Olof Wahlund }
\author[kth]{Chunliang Wang }
\author[ki,london3]{Eric Westman}
\author[]{for the Alzheimer’s Disease Neuroimaging Initiative \tnoteref{adni}}

\address[ki]{Division of Clinical Geriatrics, Department of Neurobiology, Care Sciences and Society, Karolinska Institutet, Stockholm, Sweden.}
\address[ki2]{Department of Clinical Neuroscience, Karolinska Institutet, Stockholm, Sweden.}
\address[ks]{Department of Radiology, Karolinska University Hospital, Stockholm, Sweden.}
\address[kth]{School of Technology and Health, KTH Royal Institute of Technology, Stockholm, Sweden.}
\address[london3]{Department of Neuroimaging, Centre for Neuroimaging Sciences, Institute of Psychiatry, Psychology and Neuroscience, King’s College London, London, UK.}

\cortext[mycorrespondingauthor]{Corresponding author}

\tnotetext[adni]{\tiny Data used in preparation of this article were obtained from the Alzheimer’s Disease Neuroimaging Initiative (ADNI) database (adni.loni.usc.edu). As such, the investigators within the ADNI contributed to the design and implementation of ADNI and/or provided data but did not participate in analysis or writing of this report. A complete listing of ADNI investigators can be found at: \url{http://adni.loni.usc.edu/wp-content/uploads/how_to_apply/ADNI_Acknowledgement_List.pdf}.}

\begin{abstract}
Quantifying the degree of atrophy is done clinically by neuroradiologists following established visual rating scales. For these assessments to be reliable the rater requires substantial training and experience, and even then the rating agreement between two radiologists is not perfect. We have developed a model we call \textit{AVRA} (Automatic Visual Ratings of Atrophy) based on machine learning methods and trained on 2350 visual ratings made by an experienced neuroradiologist. It provides fast and automatic ratings for Scheltens' scale of medial temporal atrophy (MTA), the frontal subscale of Pasquier's Global Cortical Atrophy (GCA-F) scale, and Koedam's scale of Posterior Atrophy (PA). We demonstrate substantial inter-rater agreement between AVRA's and a neuroradiologist ratings with Cohen's weighted kappa values of $\kappa_w$ = 0.74/0.72 (MTA left/right), $\kappa_w$ = 0.62 (GCA-F) and $\kappa_w$ = 0.74 (PA), with an inherent intra-rater agreement of $\kappa_w$ = 1. We conclude that automatic visual ratings of atrophy can potentially have great clinical and scientific value, and aim to present AVRA as a freely available toolbox. 

\end{abstract}
%\begin{keyword}
%Atrophy \sep Visual ratings \sep Machine learning \sep MRI \sep Neuroimaging
%\end{keyword}

\end{frontmatter}

%\linenumbers

%\newpage
\section{Introduction}
The assessment of structural changes in the brain is made clinically by visual ratings of brain atrophy according to established visual rating scales. They offer an efficient and inexpensive method of quantifying the degree of atrophy and can help to improve the specificity and sensitivity of dementia diagnoses \cite{Harper2015,Lars-olof2016}. However, there are limitations associated with visual ratings of atrophy, which may explain why they are still not widely used in the clinical routine. First, the ratings are inherently subjective which means that the agreement between two radiologist might be low if they have not had sufficient training \cite{Harper2015}. Second, in order to achieve adequate reliability the radiologist needs to be experienced and regularly perform ratings for the reproducibility not to drop \cite{Cavallin2012}. Third, the ratings are relatively time consuming and tedious. It takes a few minutes per image \cite{Wahlund1999}, depending on rating scale and level of rating experience. While this amount of time may be feasible in most clinical settings, it does not easily allow studying large imaging cohorts of potentially thousands of images. An automatic method would remove the inter- and intra-rater variability and eliminate the time-consuming process of rating.

\subsection{Visual rating scales}
\begin{figure}[t!]
\centerline{
\includegraphics[width=0.95\textwidth]{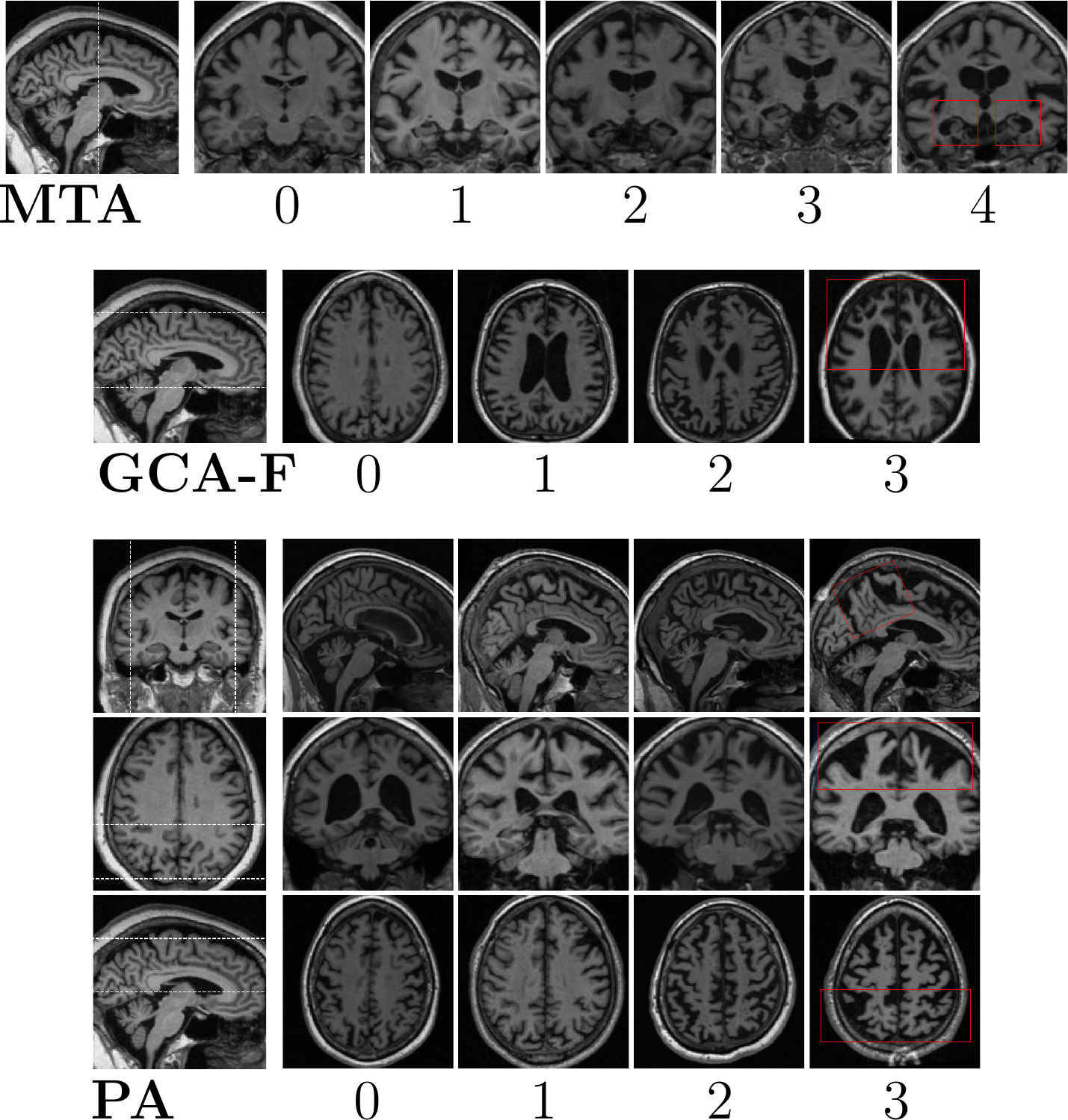}}
\caption{\label{fig:vrs} Examples of Scheltens' MTA scale \cite{Scheltens1992}, Pasquier's frontal subscale of GCA \cite{Pasquier1996}, and Koedam's PA scale \cite{Koedam2011}. The MTA ratings are done in the coronal plane, GCA-f in the axial plane, and PA ratings are based on assessments of all three planes. The area between the dashed lines in the left images indicates the slices assessed by a radiologist for the GCA-F and PA scales, while it shows the single slice assessed for MTA. The red boxes show the regions assessed for each rating scale.}
\end{figure}

Amongst the most commonly used rating scales---both in research and in clinical routine---are Scheltens' Medial Temporal Atrophy (MTA) scale \cite{Scheltens1992}, Koedam's scale for Posterior Atrophy (PA) \cite{Koedam2011} and Pasquier's scale for Global Cortical Atrophy (GCA) \cite{Scheltens1997,Pasquier1996} (see Fig. \ref{fig:vrs} for examples). These scales have previously been validated by quantitative neuroimaging techniques \cite{Bresciani2005,Henneman2009,Moller2014,Ferreira2016}. 

The MTA scale was developed by Scheltens et al. (1992) \cite{Scheltens1992}. A rating is given for each hemisphere ranging from 0 (no atrophy) to 4 (severe atrophy) and focuses on three structures: the width of the choroid fissure, the width of the temporal horn and the height of the hippocampus. The assessment is made in a single or few coronal slices on a high quality CT or ideally a T$_1$-weighted MRI. Different cut-offs have been suggested where the most common is that an average MTA score $\geq 2$ is considered pathological if the patient is younger than 75 years old, and an average MTA $\geq 3$ for patients older than 75 years \cite{Scheltens1992,Ferreira2015,Westman2011b}.

The PA scale assesses atrophy of the parietal lobe of the brain and was proposed by Koedam et al. (2011) \cite{Koedam2011}. A rating from 0 (no atrophy) to 3 (severe atrophy) is given that specifically assesses the degree of atrophy of the precuneus, the posterior cingulate sulcus, the parieto-occipital sulcus and the parietal cortex.

Pasquier et al. (1996) developed a visual rating system of cerebral atrophy in 13 different brain regions that assesses the level of dilatation of sulci and the ventricles \cite{Pasquier1996}. For each of these regions a score ranging from 0 (no atrophy) to 3 (severe atrophy) is given by the radiologist. These measures have been simplified into a global assessment of cortical atrophy rated from 0 to 3 called the GCA scale. The original paper by Pasquier and colleagues used T$_2$-weighed images\cite{Pasquier1996} but several studies have also assessed GCA in T$_1$-weighted images \cite{Ferreira2015,Ferreira2016,Ferreira2017,Scheltens1997}. A frontal subscale of GCA (GCA-F) is of particular interest since frontal atrophy has been shown to be associated with executive dysfunction \cite{Elliott2003} and can offer improved diagnosis of frontotemporal dementia (FTD) \cite{Ferreira2016}.

\subsection{Related work}
A few automatic (or semi-automatic) methods to quantify medial temporal atrophy---besides volumetrics---have previously been proposed. Two of them involve planimetrics based on manual delineation of hippocampus and surrounding structures that are combined into a single score of medial temporal atrophy \cite{Zimny2013,Menendez-Gonzalez2014}. While these methods assess almost the same structures as Scheltens' MTA scale, the different scales are not interchangeable and do not necessarily reflect the same atrophy patterns. Another study recently reported an automatic method that is trained on radiologist ratings which predicts MTA scores based on volumetric measures extracted from the MRI image \cite{Lotjonen2017}. Volumetric measures of brain regions can not be extracted from most CT images nor do they retain any information regarding the shape of the structures. It is reasonable to assume that the shapes are important since the visual MTA rating is done on a single slice, from which it is not possible to estimate the hippocampal volume.

Deep learning---a branch of machine learning---has recently generated impressive results in several fields, such as speech recognition, text semantics, image recognition and genomics \cite{LeCun2015}. Convolutional neural networks (CNN's) have already been substantially applied in medical image analysis (for recent reviews, see \cite{Shen2017a,Litjens2017}). For instance, studies using CNN's have achieved similar levels of accuracy as medical experts in classifying skin cancer \cite{Esteva2017}, mammographic skin lesion detection \cite{Kooi2017}, and diabetic retinopathy diagnosis \cite{Gulshan2016}. Focusing on applications in neuroimaging, deep neural networks have been used successfully for automatic methods of skull stripping \cite{Roy2017,Kleesiek2016}, brain age prediction \cite{Cole2017}, brain segmentation \cite{Chen2017}, PET image enhancement \cite{Wang2018} and brain tumor segmentation \cite{Pinto2016,Zhao2016} to name a few. In dementia research, several studies have investigated brains of patients with Alzheimer's disease (AD) using deep learning and shown impressive diagnostic abilities \cite{Hosseini-Asl2016,Payan2015,Suk2016,Liu2018}. A Recurrent Neural Network (RNN) is an artificial neural network that has an internal state (or "memory") and is useful when processing sequential data, such as words in a sentence or frames in a video\cite{LeCun2015,Donahue2015}. RNN's have successfully been combined with CNN's to segment MRI images, where the addition of an RNN module helped to leverage adjacent slice dependencies \cite{Ypsilantis2016,Poudel2017}.

\subsection{Our approach}
In this study, we aimed to develop an automatic algorithm based on convolutional and recurrent neural networks that provides fast, reliable, and systematic predictions of established visual ratings scales of atrophy of brain regions often affected in dementia: the MTA, GCA-F and PA scales. The models are trained on a large set of MRI images that have been rated by an experienced neuroradiologist. This method is atlas-free and requires minimum amount of setup and third-party software. We plan to present the proposed algorithm as a freely available software targeted towards neuroimaging researchers.

\section{Material and methods}
\subsection{MRI data and protocols}
Two different dementia cohorts of MRI images were included in this project: Alzheimer's Disease Neuroimaging Initiative (ADNI) and a clinical cohort with images from the memory clinic at Karolinska University Hospital (referred to as \textit{MemClin} from here on). Informed consent was obtained for all participants, or by an authorized representative of theirs.

Individuals in the MemClin cohort mainly consisted of patients clinically diagnosed with dementia according to the ICD-10 criteria between 2003 and 2011. All participants underwent a T$_1$-weighted MRI scan at the Radiology Department of Karolinska University Hospital in Stockholm, Sweden. Exclusion criteria were if the patient had other types of dementia, history of traumatic brain injury, or insufficient quality of the MRI scan \cite{Ferreira2018,Lindberg2009}. 

Data used in the preparation of this article were obtained from the Alzheimer’s Disease Neuroimaging Initiative (ADNI) database (\url{adni.loni.usc.edu}). The ADNI was launched in 2003 as a public-private partnership, led by Principal Investigator Michael W. Weiner, MD. The primary goal of ADNI has been to test whether serial magnetic resonance imaging (MRI), positron emission tomography (PET), other biological markers, and clinical and neuropsychological assessment can be combined to measure the progression of mild cognitive impairment (MCI) and early Alzheimer’s disease (AD). For up-to-date information, see \url{www.adni-info.org}. A majority of the participants in the ADNI cohort were scanned multiple times within a few weeks---often in the same day. A subset of participants were scanned both in 1.5T and 3T machines. 

All available images with an associated visual atrophy rating performed by a neuroradiologist were used in this study. Images that did not pass the initial automatic AC/PC-alignment (the anterior and posterior commissures) were excluded from the training and evaluation process (144 out of 5355 images in total). 

The algorithm was developed using theHiveDB database system\cite{Muehlboeck2014} and will become part of its automated activity system.

\subsection{Human ratings}
An experienced neuroradiologist, Lena Cavallin (L.C.), visually rated 2350 T$_1$-weighted MRI images over the course of 16 months with no prior knowledge of age, sex, or diagnosis. For ADNI subjects scanned more than once, only one of the images was rated by the radiologist and the additional image(s) were labeled with the same rating. The distribution of L.C.'s MTA, PA and GCA-F ratings are shown in Table \ref{tab:ratings}. Many of the ADNI ratings have been analyzed and reported in previous studies \cite{Ferreira2018,Ferreira2015,Ferreira2016,Ferreira2017,Westman2011b}. All visual ratings of MTA, PA and GCA-F were based on T$_1$-weighted MRI images, and illustrative examples of the ratings can be seen in Fig. \ref{fig:vrs}. The images were aligned with AC/PC by the radiologist if the protocol allowed for it \cite{Cavallin2012}. The MTA ratings were made in a single coronal slice, just behind the amygdala and mammillary bodies. The GCA-F ratings were based on multiple sagittal slices, whereas the PA score was based on slices in all three planes.

\begin{table}[t!]
\caption{\label{tab:ratings} The rating distribution of the images used in the study. The "Images" column refers to how many \textit{unique} images that were rated by the radiologist at least once. Both the left and right MTA ratings are presented in the "MTA" column in the Table.}
\small
\centerline{
\begin{tabular}{|l|c|c c c c c|c c c c|c c c c|}
\hline
\multirow{2}{*}{\textbf{Cohort}} & \multirow{2}{*}{\textbf{Images}} & \multicolumn{5}{c|}{\textbf{MTA}}  & \multicolumn{4}{c|}{\textbf{GCA-F}} & \multicolumn{4}{c|}{\textbf{PA}} \\
\cline{3-15}
& & 0 & 1& 2& 3& 4 & 0 & 1 & 2 & 3 & 0 & 1 & 2 & 3 \\\hline
ADNI & 1966 & 425 & 1581 & 1147 & 555 & 224 & 1449 & 468 & 49 & 0 & 1188 & 611 & 157 & 10\\\hline
MemClin & 384 & 23 & 265 & 296 & 139 & 45 & 279 & 89 & 14 & 2 & 210 & 127 & 43 & 4\\\hline
\hline Total & 2350 & 448 & 1846 & 1443 & 694 & 269 & 1728 & 557 & 63 & 2 & 1398 & 738 & 200 & 14\\\hline
\end{tabular}}
\end{table}

\begin{figure}
\centerline{
\includegraphics[width=.8\textwidth]{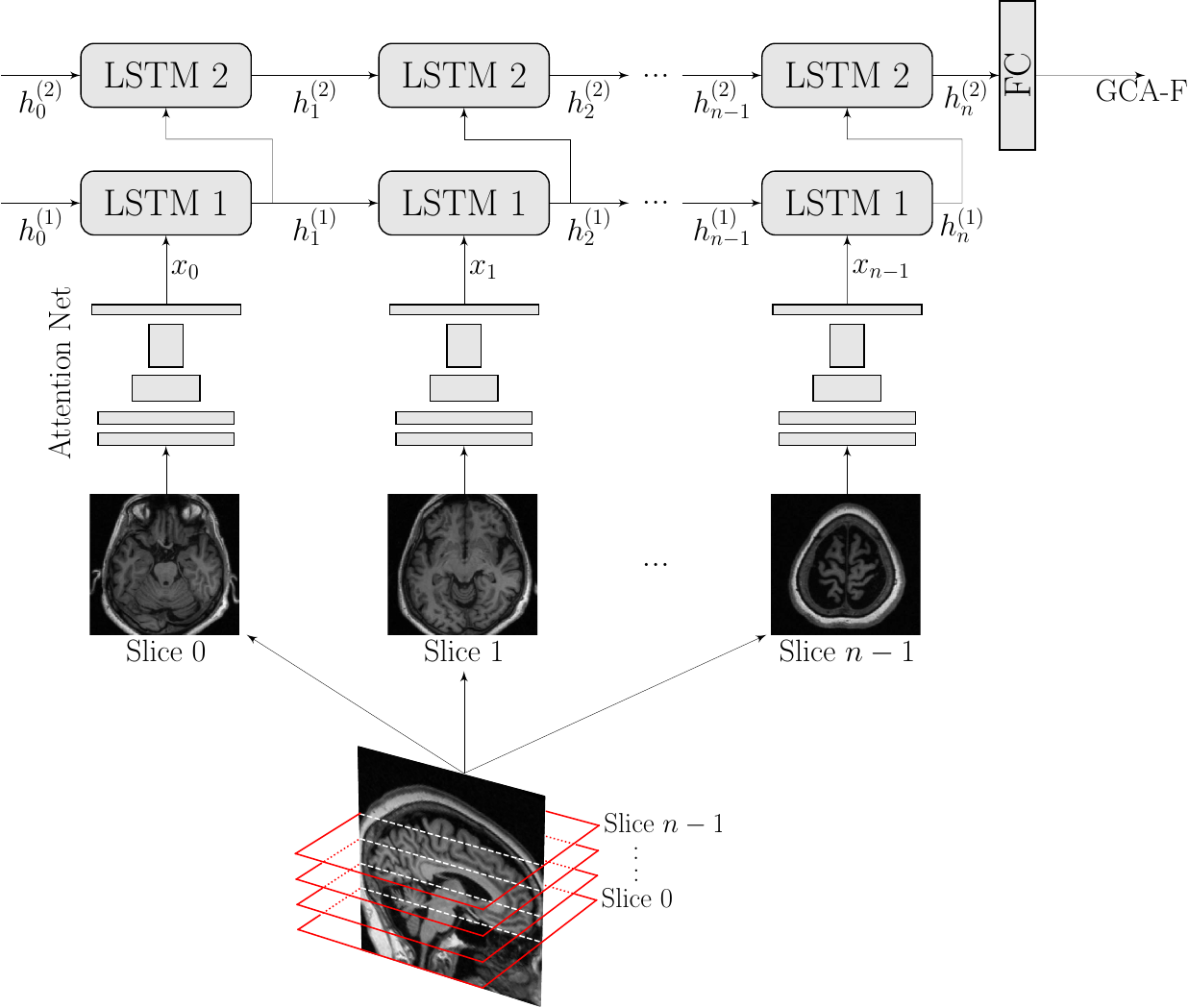}}
\caption{\label{fig:avra} A sketch of the architecture of AVRA, with the example of a GCA-F prediction. The MTA and PA models followed the same structure.}
\end{figure}
\subsection{Computer ratings}
The motivation behind the proposed model architecture was to mimic how a neuroradiologist would process an MRI image: to scroll through the brain volume slice-by-slice looking for the "correct" slice(s) to base the rating on. A human rater assesses images acquired using different scanners, vendors and protocols without any need for substantial preprocessing such as segmentation, intensity normalization, non-linear registrations or skull-stripping. To better mimic the clinical situation (and to keep the number of time consuming preprocessing steps that can potentially fail to a minimum) we trained AVRA to rate images with as little preprocessing as possible. The main difference between AVRA's and a human rater is that AVRA's ratings are continuous instead of discrete.

All code in this project was developed in Python 3.4.3 using the deep learning framework PyTorch 1.0 \cite{Paszke2017}.

\subsubsection*{Preprocessing}
The only preprocessing included in our method is the registration of all brains to the MNI standard brain using FSL FLIRT 6.0 (FMRIB's Linear Image Registration Tool) \cite{Jenkinson2002,Jenkinson2001,Greve2009}. This rigid transform is computed with 6 degrees of freedom (i.e. rotation and translation only) and is used to automatically AC-PC align each brain and conform all images to the same voxel size (1x1x1mm$^3$) and input dimension (182x218x182). The AC-PC aligned images are cropped to remove excess space outside the brain and redundant slices not part of the ratings scale (as indicated in Figs. \ref{fig:vrs} and \ref{fig:avra}). The center-voxel of the cropped images depended on the rating scale. For the MTA ratings, 22 coronal slices of the dimension 128mm x 128mm are input to the model---enough to ensure that the "correct" rating slice is included. The GCA-F ratings are done on multiple axial slices so each volume is cropped to 160mm x 192mm x 40 slices, with 2mm slice thickness. The PA model requires slices from all three anatomical planes. From each MRI image a smaller volume of 128mm x 128mm x 128mm was extracted from the parietal lobe, sufficiently large to include all relevant structures in the parietal cortex. From this cropped volume 37 axial, 28 coronal and 34 sagittal slices with 2mm slice thickness (i.e. 99 slices in total) were used as input to the model. Since the distribution of raw voxel values was very different---particularly between 1.5T and 3T images---all cropped volumetric images were normalized to have a zero variance and mean.

\subsubsection*{Model architectures}
\begin{figure}[t!]
\centerline{
\includegraphics[width=1\textwidth]{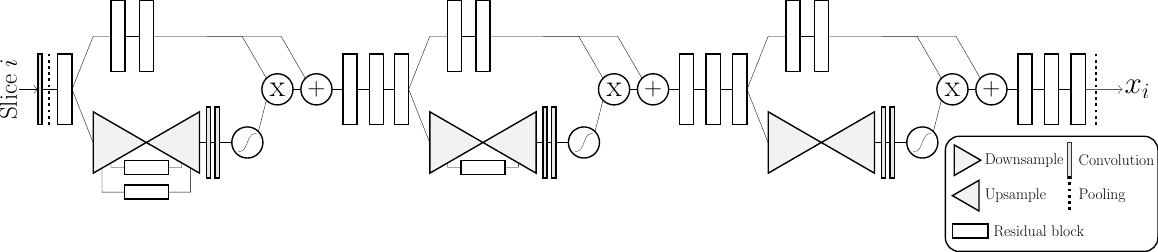}}
\caption{\label{fig:resattnet} A sketch of the residual attention net used to extract features from individual MRI slices, where the flattened output is fed to the RNN. The downsampling block consists of stacking maxpooling operations followed by a residual block. The upsamling is performed with bilinear interpolations of the output of a residual block. The "+", "x", and "S-shaped" symbols denote element-wise summation, multiplication and the sigmoid function, respectively. The flow chart is adapted from \cite{Wang2017}.}
\end{figure}

The overall structure of the models can be seen in Fig. \ref{fig:avra} and can be split into three parts. First, relevant features from a single slice are extracted using a Residual Attention Network \cite{Wang2017}, detailed in Fig. \ref{fig:resattnet}. It combines the abilities from residual learning \cite{He2015}, which can allow for even deeper models, and attention models that can "focus" spatially on images---particularly useful for visual ratings since they are based on regional atrophy \cite{Xu2015,Ba2015}. Our implementation is a slimmed version of the original, with the same depth but a smaller number of filters in each layer to reduce memory usage and computation time. Initial experiments showed no noticeable performance reduction on the validation set compared to using a larger network. Second, the features are reshaped to a 1D vector and fed to an RNN, which consists of a two-layer Long-Short Term Memory (LSTM) network with 256 hidden nodes \cite{Hochreiter1997,Gers2000}. The LSTM modules are expected to "remember" relevant features seen in previous slices and update its state ("memory") when it is exposed to a slice containing useful information for the rating. Finally, when slice 0, 1, ..., $(n-1)$ have been propagated through the network, the final output from the second LSTM module $h_n^{(2)}$ is used to make a linear prediction of the visual rating. All three models share the same network architecture except for the size of the input vector fed to the LSTM network, as that is dependent on the input size of the MRI slices.

For comparison, we train a VGG16 network \cite{Simonyan2015} without the RNN part, where the 3D volumes are treated as multi-channel 2D images. That is, for the MTA model we input one "22-channel" image to the CNN once instead of 22 single-slice images.

\subsubsection*{Training}
For training and evaluation, the dataset was randomly split into a training and a hold-out test set, where 20\% of all subjects where assigned to the test set. On the remaining images in the training set we applied 5-fold cross validation for hyper-parameter tuning for each rating scale. The five trained models were used together as an ensemble classifier evaluated on the test set, where the average prediction was considered the final rating.

The models were trained for 200 epochs using backpropagation and optimized through stochastic gradient descent (SGD) with cyclic learning rate to maximize the probability of predicting the radiologist's rating \cite{Loshchilov2016,Huang2017}. The training set was randomly split into minibatches, each containing 20 MRI images, and the weights were updated to minimize the mean-squared error between the automatic and the integer ratings by L.C. We employed data augmentation in the training process of the network to reduce the risk of overfitting to the training set. This included random cropping (within $\pm$10mm off the center voxel), scaling, left/right mirroring, and randomly selecting N4ITK inhomogeniety corrected images instead of the original file \cite{Tustison2010}. Due to the imbalance of ratings in the dataset we employed random oversampling of images with less frequent ratings, which has been shown to improve the prediction performance of CNN's \cite{Buda2017}. For ADNI subjects that had multiple scans for a single timepoint, a scan was selected randomly for each minibatch.

\subsection{Analyses metrics}
The visual rating scales are subjective measures by definition. Consequently, there are no objective ground truth ratings available. In most studies, the performance of a rater is reported in kappa statistics---a group of measures that can quantify the level of agreement between two sets of discrete ratings---but there is no single metric always reported. To make our results comparable to previous findings, we present our results with Cohen's weighted kappa ($\kappa_w$), which has been used in several previous rating studies \cite{Koedam2011,Westman2011b,Cavallin2012,Cavallin2012a,Ferreira2016,Ferreira2017,Velickaite2017}, as well as accuracy and the Pearson correlation coefficient ($\rho$). The agreement between two sets of ratings is referred to \textit{inter}-rater agreement if the sets were assessed by different raters, and \textit{intra}-rater agreement if a single radiologist rated the set twice. 

\section{Results}
\subsection{Intra-rater agreements}
To have an idea of the variability in the human ratings used for training in this project, we studied the intra-rater agreement in a subset of 244 images that had been rated 2-4 times with at most 16 months from the first to the last rating session. To be consistent with the computer training and evaluation procedure, we compared the latest rating to a previous one. If there were more than two ratings, the previous rating was chosen randomly. This yielded $\kappa_w$ agreements and accuracies for MTA (left): $\kappa_w$ = 0.83, acc = 76\%; MTA (right): $\kappa_w$ = 0.79, acc = 70\%; GCA-F: $\kappa_w$ = 0.46, acc = 71\%; PA $\kappa_w$ = 0.65, acc = 72\%. Ratings made only 1 week apart showed substantially better intra-rater agreement (see \textit{Ferreira et al. (2017)} entry in Table \ref{tab:rater}). These results provide an estimate of the "human-level agreement"---i.e. approximate levels of agreement our models should be able to achieve by training on the available cohort due to rating inconsistencies over 16 months.

Since there are no random elements in the evaluation process of a brain image, the "intra-rater" agreement of AVRA is inherently $\kappa_w = 1$.

\subsection{Inter-rater agreements}
Our models predicted continuous rating scores of an image, based on training from discrete ratings by L.C. We rounded AVRA's ratings to the nearest integer to be able to compare the rating consensus in terms of accuracy and kappa statistics. The agreements between the radiologist's and AVRA's (as well as the VGG networks') ratings on the hold-out test set are summarized in Table \ref{tab:rater} together with previously reported $\kappa_w$ values of inter- and intra-rater agreements. The inter-rater agreement $k_w$, Pearson correlation $\rho$, and accuracy on the test set for MTA (left): $\kappa_w$ = 0.74, acc = 70 \%; MTA (right): $\kappa_w$ = 0.72, $\rho$ = 0.88, acc = 70 \%; GCA-F: $\kappa_w$ = 0.62, $\rho$ = 0.71, acc = 84 \%; PA: $\kappa_w$ = 0.74, $\rho$ = 0.85, acc = 83\%. These agreement levels were similar to previously reported in studies, see Table \ref{tab:rater}. The naive VGG16 implementations showed lower inter-rater agreements with the radiologist compared to AVRA.

\begin{table}
\caption{\label{tab:rater} Previously reported intra- and inter-rater agreements together with the test set agreement between L.C. and AVRA, and L.C and VGG16 as a reference. The interval given refers to the minimum and maximum weighted kappa ($\kappa_w$) value reported in the referenced study.}
\centerline{
\small
\begin{tabular}{|l|c|c|c|c|}
\hline
\textbf{\thead{Study}} & \textbf{\thead{Scale}} & \textbf{\thead{$N$}} & \textbf{\thead{Intra-rater\\ agreement ($\kappa_w$)}} & \textbf{\thead{Inter-rater\\ agreement ($\kappa_w$)}} \\\hline 
Cavallin et al. (2012) \cite{Cavallin2012} & MTA & 100 & 0.83-0.94 &  0.72 - 0.84 \\
Cavallin et al. (2012b) \cite{Cavallin2012a} & MTA & 100 &  0.84-0.85  &  --- \\
Westman et al. (2011) \cite{Westman2011b} & MTA & 100 & 0.93  &  --- \\
Velickaite et al. (2017) \cite{Velickaite2017} & MTA & 390 & 0.79-0.84   &   0.6-0.65  \\
Ferreira et al. (2017) \cite{Ferreira2017} & MTA & 120 & 0.89-0.94  &  0.70-0.71  \\
Koedam et al. (2016) \cite{Koedam2011} & MTA & 29-118 & 0.91-0.95  & 0.82-0.90  \\
VGG16 & MTA & 464 & 1  & 0.58 - 0.59 \\
\textbf{AVRA} & \textbf{MTA} & \textbf{464} &  1 & \textbf{0.72 - 0.74}   \\\hline
Koedam et al. (2016) \cite{Koedam2011} & PA & 29-118 & 0.93-0.95  & 0.65-0.84  \\
Ferreira et al. (2017) \cite{Ferreira2017} & PA & 120 & 0.88  &  0.88  \\
VGG16 & PA & 464 & 1  & 0.63  \\
\textbf{AVRA} & \textbf{PA} & \textbf{464} &\textbf{1} &   \textbf{0.74} \\\hline
Ferreira et al. (2016) \cite{Ferreira2016} & GCA-F & 100 & 0.70  &  0.59  \\
Ferreira et al. (2017) \cite{Ferreira2017} & GCA-F & 120 & 0.83  &  0.79  \\
VGG16 & GCA-F & 464 & 1  & 0.56  \\
\textbf{AVRA} & \textbf{GCA-f} & \textbf{464} &  \textbf{1} &  \textbf{0.62} \\\hline
\end{tabular}}
\end{table}

\begin{figure}[t!]
\centerline{
\includegraphics[height=0.18\textheight]{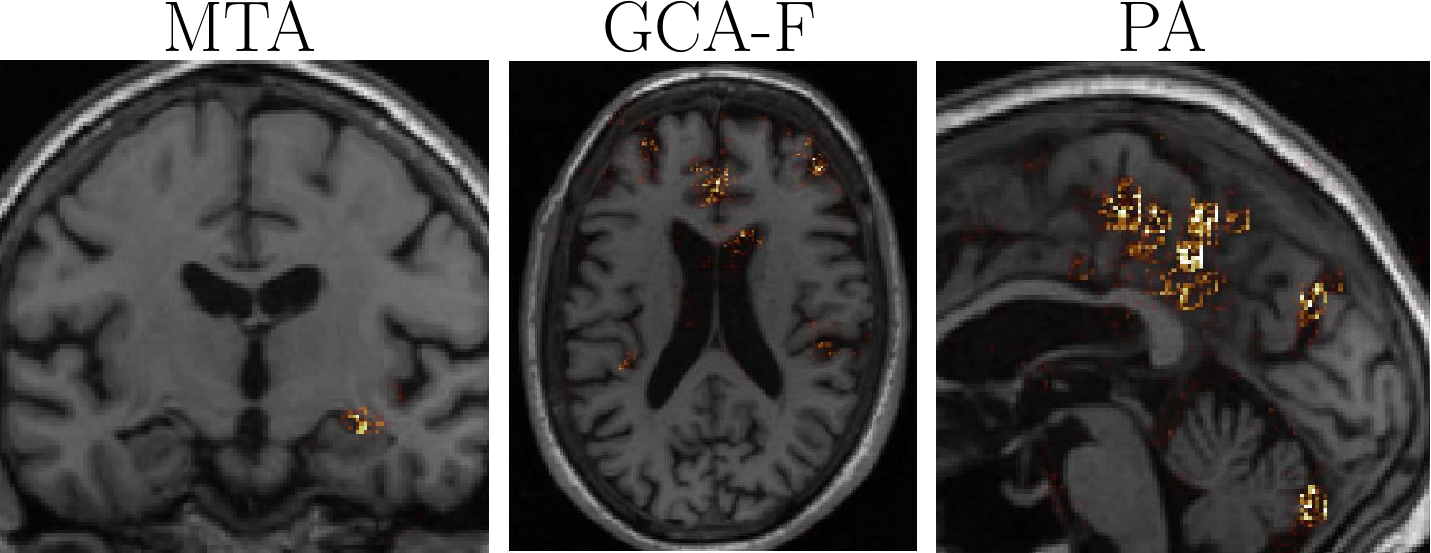}
}
\caption{\label{fig:sal} Examples of sensitivity maps for the MTA, GCA-F and PA scale, respectively. These maps indicate the influence each voxel had in AVRA's rating. The particular slices displayed were chosen manually as representative images for each rating scale.}
\end{figure}

To increase interpretability and understanding of the models, we computed gradient-based sensitivity maps of images in the test set based on the SmoothGrad method \cite{Smilkov2017}. These indicated how influential individual voxels were in the rating prediction, which we can apply to verify that the network identified the correct features. Examples of AVRA's rating predictions for each scale are shown in Fig. \ref{fig:sal}. As can be observed, the MTA sensitivity maps were generally focused only around the area of the hippocampus and the inferior lateral ventricle in $\sim \pm$ 3 slices from the "correct" rating slice. The sensitivity maps in other more posterior and anterior slices were close to zero. The GCA-F maps were more diffused, but the greatest magnitudes were primarily seen in the sulci of the frontal lobe. The PA maps were mainly visible in the parietal lobe and in the sagittal plane, with the greatest magnitudes appearing in parieto-occiptal sulcus and precuneus. 

\section{Discussion}
We have developed a tool for automatic visual ratings of atrophy (AVRA) that is fast, systematic and robust. AVRA is trained on a large set of images rated by an expert neuroradiologist using the established clinical assessment measures of Scheltens' MTA scale, Pasquier's GCA-F scale and Koedam's PA scale with agreement levels similar to that between two experienced radiologists. This tool runs in under 1 minute on a regular laptop, which enables automatically rating thousands of images in a couple of hours. Rating an MRI image of the brain requires minimum amount of preprocessing and the models were built to potentially work in a clinical setting. The main advantage of an automatic model is the absence of randomness, which can ensure rating consistency between different clinics, research groups and cohorts. Thus, AVRA has potential to function as a clinical aid, and to increase the use of visual ratings in research.

\subsection{Agreement levels}
The rating agreements between AVRA's and the radiologist's ratings were considered \textit{substantial} (i.e. between 0.6-0.8) according to the often cited paper by Landis and Koch (1977) \cite{Landis1977}. The agreements were close to the "human-level agreements" in this study (i.e. the agreement between the multiple L.C. ratings of the same image). This was reasonable since a model trained on imperfect labels due to rating inconsistency can never achieve perfect agreement. A previous study has investigated the overtime reliability of MTA ratings, where their results showed that the intra-rater agreement is typically higher when a set is rated twice closer in time--especially when the radiologist do not rate images on a daily basis \cite{Cavallin2012}. The time between ratings is often not reported, but in Pasquier's introduction of the GCA scale the second rating was performed 24 h after the first \cite{Pasquier1996}. Thus it is reasonable that if all images in a study were rated twice 16 months apart, the intra-rater agreement would generally be lower than the actual reported values. Our analysis of the subset of images rated more than once suggests this to be the case. Those values may not necessarily reflect the "true" rating consistency either since the multiple-rating subset does not follow the same distribution as the whole cohort. Limiting the time span between the first and last set of ratings meant having to discard a large part of the images in the training set, and initial investigations of this showed decreasing agreement in our study. This suggested that a large number of images for training was more important than the potential inclusion of noisy labels.

AVRA's ratings agreed more with the radiologist ratings than the VGG16 models' did. A recurrent CNN architecture might thus be particularly suitable for visual rating predictions, but we can not say from these results if it were the residual modules, the attention components, or the LSTM cells---all used in AVRA but not in the VGG16 models---that had the greatest positive impact on the performance. Another contributing factor may be the wide difference in the number of trainable parameters between AVRA (1.5M) and VGG16 (65M) that makes AVRA less prone to overfit on the training data. However, it should be noted that we spent more time to tune and optimize AVRA compared to the VGG16 networks, which biases the results in favor of AVRA. 

The automatic model presented by Lötjönen and colleagues (2017) is, to our knowledge, the only software that also attempts to predict scores based on clinical visual rating scales \cite{Lotjonen2017}. It is based on volume measures of hippocampus and surrounding structures, whereas AVRA predicts the ratings directly from the voxel intensity values. This makes our proposed method promising to also work on MRI images with large slice thickness and CT images, from which volumes generally cannot be computed. The fact that CT is a cheaper and more commonly used imaging modality than MRI in the clinics speaks in favor of using convolutional neural networks over volumetrics for automatic ratings of atrophy \cite{Falahati2015}. No $\kappa_w$ values are reported in \cite{Lotjonen2017}, but they provided correlation coefficients between radiologist and computer ratings for the MTA scale as 0.86 (left) and 0.85 (right). AVRA showed a similar magnitude of correlation for the MTA scale on the hold-out test set: $\rho= 0.88$.

\begin{figure}[t!]
\centerline{
\includegraphics[width=1\textwidth]{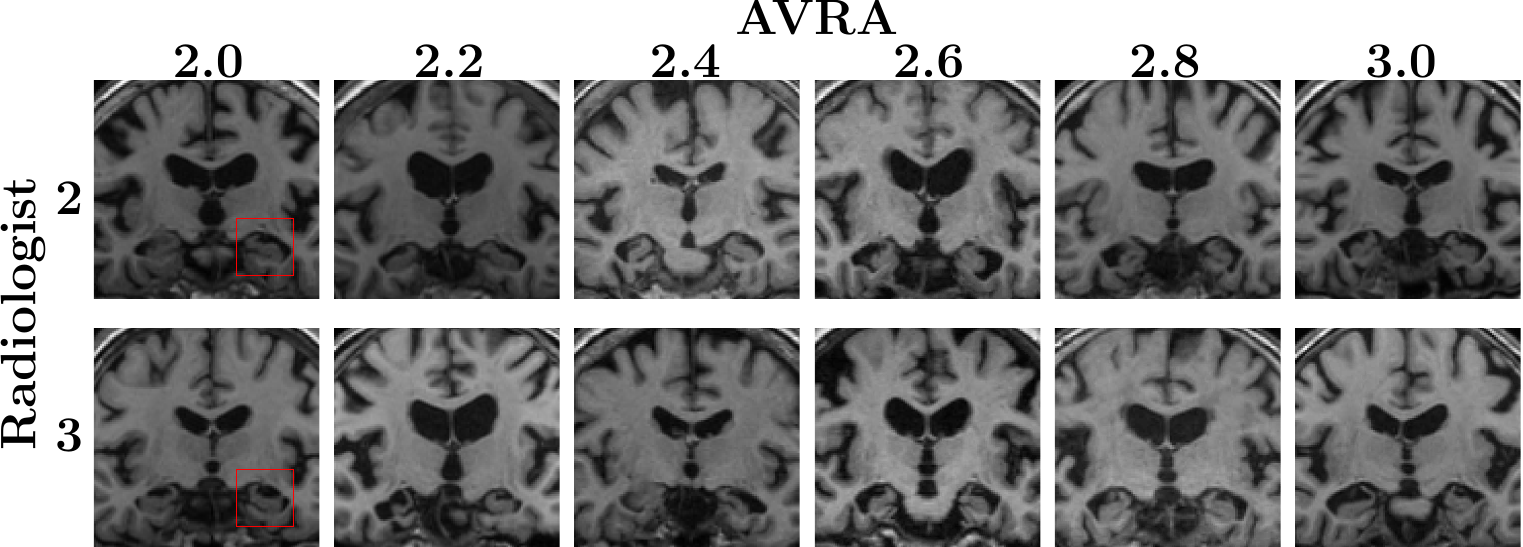}}
\caption{\label{fig:miss} Comparison between AVRA's continuous ratings and the neuroradiologist's discrete ratings of the same images. \textit{Rows}: MRI slices with MTA on the right side of the image (side indicated by the red squares) rating of 2 (top) and 3 (bottom) given by neuroradiologist. \textit{Columns}: corresponding continuous AVRA ratings. E.g., the second image from the left in the bottom row was given assessed to have a left MTA score of 3 by the neuroradiologist and 2.2 by AVRA. When the radiologist re-examined these cases the same ratings were given for all images, except for the three images on the right in the top row (Radiologist: 2, AVRA: 2.6, 2.8 and 3.0), which were instead given MTA scores of 3. The image rated 2 by L.C. and 2.4 by AVRA was described as a subject between 2 and 3.}
\end{figure}

\subsection{Reliability of AVRA}
One of the main motivations of having a computer rate brain atrophy instead of humans is its inherent perfect intra-rater agreement---the same image will be rated exactly the same regardless of when (and where) it is rated. A relevant question to ask is: why not let a computer segment and calculate e.g. hippocampal volumes instead of an MTA rating? We see three main motivations for this: 1) CT, and some MRI protocols, have too large slice thickness that do not allow for extracting reliable volumetric information from the images. 2) Segmentaion methods will---just as AVRA---fail in processing some cases, and for clinician to manually intervene and delineate structures would neither be feasible nor practical. If an automatic visual rating would fail the radiologist would be able to quickly perform their own visual rating, as is done today. 3) There is a lack of how to clinically interpret volumetric data, e.g. the hippocampal volumes. However, extensive research has been done on cut-offs for visual rating scales, even considering modulating factors such as age \cite{Ferreira2015}.

The sensitivity maps shown in Fig. \ref{fig:sal} suggested that the models were able to correctly identify relevant structures to base their ratings on. Particularly the sensitivity maps of the MTA model were typically not visible $\pm3$mm from the "correct" rating slice, indicating that the employed recurrent CNN architecture used was able to correctly identify relevant slices and disregard redundant ones. The diffused sensitivity maps seen for the GCA-F scale was also observed in the quantitative validation study done by Ferreira et al. (2015), showing that frontal atrophy is also associated with temporal and posterior atrophy---at least in the ADNI cohort \cite{Ferreira2016}. Möller and colleagues (2014) found, using VBM analysis, significant differences between PA ratings not only in the parietal lobe, but also in parts of the cerebellum, temporal lobe and the occipital lobe \cite{Moller2014}. Their study was also performed on a cohort with individuals with probable AD and subjective memory complaints, concluding that atrophy solely in the posterior cortex is an exception. The sensitivity maps from our PA model indicate that AVRA based the PA ratings on mainly the same regions. AVRA learns to how to predict a GCA-F or a PA score from an MRI image \textit{only} based on previous human ratings. Thus, if e.g. frontal atrophy is strongly associated with atrophy in the temporal lobe, the model is likely to find it difficult to learn to only assess the frontal lobe in the GCA-F scale. Since the sensitivity maps are based on the absolute values of the calculated gradients in the backward propagation, the magnitude of these decrease every time it propagates through the LSTM cell due to the point-wise multiplication in the forget gate \cite{Gers2000}. The PA model inputs 99 slices. As the sagittal slices are the last to be fed to the model it is reasonable to assume that they dominate the sensitivity maps as opposed to early axial slices, which have propagated through the LSTM cell almost 100 times.

The performance of AVRA was validated in a test set that was randomly sampled from the same cohorts as the training data set. This means that the data distribution in the test set was similar to the image samples that the models were trained on. This is a simpler test set than if the test set was from a different cohort with images acquired using other scanning parameters. We are currently in the process of validating how the models would handle data from a different image distribution (cohort), and the effect it would have on the rating agreement.

Frequently, it is difficult for a radiologist to decide between two scores, and in a clinical situation the level of atrophy is often described as "the left MTA is between 2 and 3" for instance. This nuance might be important information for the physician diagnosing dementia, but in research single integer scores have typically been used following the original definitions of the rating scales. Previous attempts of (semi-)automatic atrophy measures have output a continuous measure \cite{Zimny2013,Menendez-Gonzalez2014,MenendezGonzalez2016,Lotjonen2017}. The main advantages of using a continuous measure of atrophy are 1) atrophy evolves continuously and thus it is reasonable to describe its degree through a continuous measure, and 2) it provides more detailed information about the severity of the atrophy. The latter point is for instance particularly useful to track disease progression and could allow us to establish more sensitive cut-off values for different diagnoses. It is also easy to convert the continuous measures of the rating scales to their discrete, original versions by rounding to nearest integer. 

In Fig. \ref{fig:miss} we show some examples between AVRA's continuous and the radiologist discrete ratings in the important diagnostic interval between MTA=2 and MTA=3. When studying these images again post AVRA's ratings, the radiologist only assessed that the images originally rated MTA=2 with associated AVRA scores of 2.6-3.0 to be wrongly rated. They would be re-rated as MTA=3, i.e. closer to AVRA's score. The image scored MTA=2 (radiologist) and MTA=2.4 (AVRA) was described as a case between 2 and 3, which may illustrate the usefulness of continuous ratings. However, we noticed that in two of the most disagreeing ratings (L.C.: MTA=3, Avra: MTA=\{2.0, 2.2\}) the individuals had an adhesion between the hippocampus and the cerebral white matter. These cases are not frequent, and the rating disagreements in Fig. \ref{fig:miss} indicate that AVRA did not learn to correctly adjust the score for the presence of adhesions.

\begin{figure}
\centerline{
\includegraphics[trim={0.5cm 0cm 2cm 1cm},clip=true,width=0.7\textwidth]{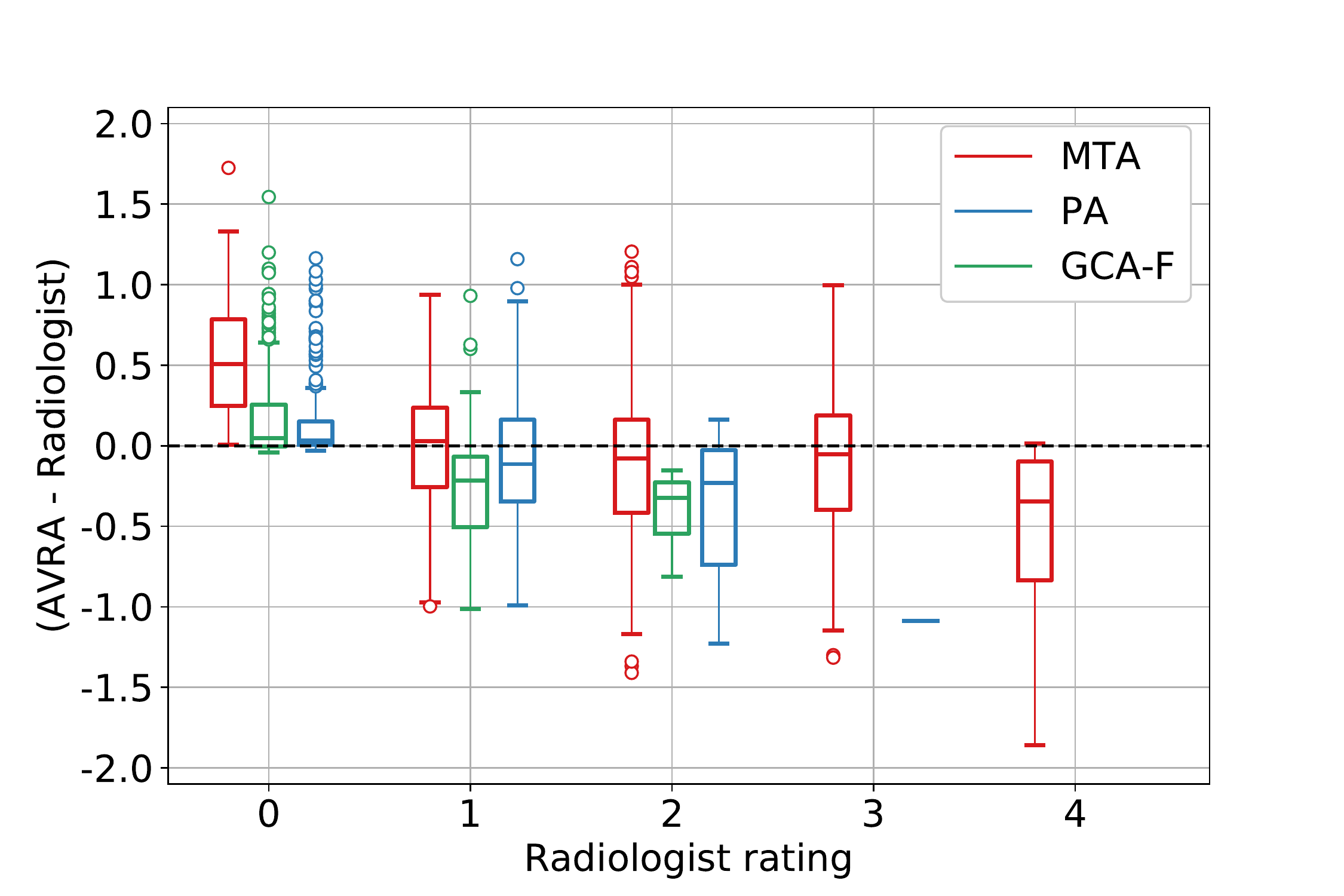}
}
\caption{\label{fig:dif} Box plots of the difference between AVRA's continuous and the radiologist's discrete ratings of the same image (stratified by radiologist score) for the MTA (red), GCA-F (green), and the PA scale (blue). There were no images assigned a rating of GCA-F=3 by the radiologist and only 1 image with PA=3 in the test set, which explains the absence of boxes for these ratings.}
\end{figure}

We aimed to design AVRA to function on images with the least amount of preprocessing possible to demonstrate that it could work in a clinical setting. A few concessions were made to facilitate the training process---mainly the AC-PC alignment performed through rigid registration to the MNI brain using FSL FLIRT. This helped centering all images to allow for tighter cropping around the structures of interest. However, this automatic preprocessing step failed in around 2.5\% of all images, which were discarded for future training and evaluation although the quality of most of the images was good enough for a radiologist to visually rate it. Since the MRI image input to CNN has not been intensity normalized, skull stripped or motion corrected, it is possible to perform a manual AC-PC alignment for the failed cases and then input them to the model. More extensive data augmentation and training, or using reinforcement learning to find the correct slice, could potentially be used to avoid the AC-PC alignment step and just input the raw MRI image. This was, however, not explored in the current study. 

\subsection{Limitations}
There are some limitations of the proposed algorithm. First, the models are solely based on the ratings by a single radiologist and thus assume that the ratings we trained the model on are "ground truth" labels. A model trained on these labels can therefore never be "better" than the rater. If the ratings have systematic errors the model will incorporate these. For instance, a rater might systematically look at the left medial temporal lobe when rating the MTA of the right hemisphere, which could influence (bias) the right hemisphere MTA score. If we train a model on these ratings, this bias would be learned by the model as well. Another approach would be to have multiple expert radiologists rate a set of images together or separately and use these labels as ground truth. However, it is not feasible to have multiple radiologist visually assess the large number of images necessary for training a deep neural network. It also does not automatically mean that these ratings would necessarily be "closer" to the ground truth. If future studies want to use a neural network based on their own set of ratings, it should be possible to start from the pre-trained networks of AVRA and fine-tune the final classification layer(s) on the new ratings. This would require substantially fewer ratings, since the convolutional part would already have learned to extract relevant features from the images.

The second limitation of the study are the small numbers of the highest GCA-F and PA ratings, which may increase the risk of "true" 3 score to be misclassified. Based on the results in Fig. \ref{fig:dif} this seems to be the case. As the diagnostic cut-off values for these ratings scales in AD diagnosis have been suggested as PA $\geq$1 and GCA-F$\geq$1 \cite{Ferreira2015}, the clinical implications of this may be minor even in the cases where the atrophy is rated as a 2 instead of a 3. These severe ratings are rare also in previous studies on dementia cohorts \cite{Ferreira2015,Rhodius-Meester2017}, so this will likely be an issue for any computerized method trained on radiologist ratings.

\section{Conclusion}
In this study, we have proposed an automatic method (AVRA) to provide visual ratings of atrophy according to Scheltens' MTA scale, Koedam's PA scale, and Pasquier's frontal GCA scale. AVRA mimics the neuroradiologist's rating procedure and achieves similar levels of agreement to that between two experienced neuroradiologists---without any prior preprocessing of the MRI images. We plan to make AVRA freely available as a user-friendly software aimed towards neuroscientists and neuroradiologists.

\section*{Acknowledgements}
We would like to thank the Swedish Foundation for Strategic Research (SSF), The Swedish Research Council (VR), the Strategic Research Programme in Neuroscience at Karolinska Institutet (StratNeuro), Swedish Brain Power, the regional agreement on medical training and clinical research (ALF) between Stockholm County Council and Karolinska Institutet, Hjärnfonden, Alzheimerfonden, the Åke Wiberg Foundation and Birgitta och Sten Westerberg for additional financial support.

Data collection and sharing for this project was funded by the Alzheimer's Disease Neuroimaging Initiative (ADNI) (National Institutes of Health Grant U01 AG024904) and DOD ADNI (Department of Defense award number W81XWH-12-2-0012). ADNI is funded by the National Institute on Aging, the National Institute of Biomedical Imaging and Bioengineering, and through generous contributions from the following: AbbVie, Alzheimer’s Association; Alzheimer’s Drug Discovery Foundation; Araclon Biotech; BioClinica, Inc.; Biogen; Bristol-Myers Squibb Company; CereSpir, Inc.; Cogstate; Eisai Inc.; Elan Pharmaceuticals, Inc.; Eli Lilly and Company; EuroImmun; F. Hoffmann-La Roche Ltd and its affiliated company Genentech, Inc.; Fujirebio; GE Healthcare; IXICO Ltd.; Janssen Alzheimer Immunotherapy Research \& Development, LLC.; Johnson \& Johnson Pharmaceutical Research \& Development LLC.; Lumosity; Lundbeck; Merck \& Co., Inc.; Meso Scale Diagnostics, LLC.; NeuroRx Research; Neurotrack Technologies; Novartis Pharmaceuticals Corporation; Pfizer Inc.; Piramal Imaging; Servier; Takeda Pharmaceutical Company; and Transition Therapeutics. The Canadian Institutes of Health Research is providing funds to support ADNI clinical sites in Canada. Private sector contributions are facilitated by the Foundation for the National Institutes of Health (\url{www.fnih.org}). The grantee organization is the Northern California Institute for Research and Education, and the study is coordinated by the Alzheimer’s Therapeutic Research Institute at the University of Southern California. ADNI data are disseminated by the Laboratory for Neuro Imaging at the University of Southern California.

%\section*{References}
\markboth{References}{References}
\renewcommand\refname{References}
\small
\bibliography{references}{}

\begin{thebibliography}{10}

\bibitem{Harper2015}
Lorna Harper, Frederik Barkhof, Nick~C. Fox, and Jonathan~M. Schott.
\newblock {Using visual rating to diagnose dementia: A critical evaluation of
  MRI atrophy scales}.
\newblock {\em Journal of Neurology, Neurosurgery and Psychiatry},
  86(11):1225--1233, 2015.

\bibitem{Lars-olof2016}
Lars~Olof Wahlund, Eric Westman, Danielle van Westen, Anders Wallin, Sara
  Shams, Lena Cavallin, and Elna~Marie Larsson.
\newblock {Imaging biomarkers of dementia: recommended visual rating scales
  with teaching cases}.
\newblock {\em Insights into Imaging}, 8(1):79--90, 2017.

\bibitem{Cavallin2012}
Lena Cavallin, Kirsti L{\o}ken, Knut Engedal, Anne~Rita {\O}kseng{\aa}rd,
  Lars~Olof Wahlund, Lena Bronge, and Rimma Axelsson.
\newblock {Overtime reliability of medial temporal lobe atrophy rating in a
  clinical setting}.
\newblock {\em Acta Radiologica}, 53(3):318--323, 2012.

\bibitem{Wahlund1999}
Lars-Olof Wahlund, Per Julin, Johan Lindqvist, and Philip Scheltens.
\newblock {Visual assessment of medial temporal lobe atrophy in demented and
  healthy control subjects: correlation with volumetry}.
\newblock {\em Psychiatry Research: Neuroimaging}, 90(3):193--199, 1999.

\bibitem{Scheltens1992}
Philip Scheltens, D~Leys, F~Barkhof, D~Huglo, H~C Weinstein, P~Vermersch,
  M~Kuiper, M~Steinling, E~Ch Wolters, and J~Valk.
\newblock {Atrophy of medial temporal lobes on MRI in "probable" Alzheimer's
  disease and normal ageing: diagnostic value and neuropsychological
  correlates}.
\newblock {\em Journal of Neurology Neurosurgery, and Psychiatry}, 55:967--972,
  1992.

\bibitem{Koedam2011}
Esther~L.G.E. Koedam, Manja Lehmann, Wiesje~M. {Van Der Flier}, Philip
  Scheltens, Yolande~A.L. Pijnenburg, Nick Fox, Frederik Barkhof, and Mike~P.
  Wattjes.
\newblock {Visual assessment of posterior atrophy development of a MRI rating
  scale}.
\newblock {\em European Radiology}, 21(12):2618--2625, 2011.

\bibitem{Scheltens1997}
Philip Scheltens, Florence Pasquier, Jan~G.E. Weerts, Frederik Barkhof, and
  Didier Leys.
\newblock {Qualitative assessment of cerebral atrophy on MRI: inter- and intra-
  observer reproducibility in dementia and normal aging}.
\newblock {\em European Neurology}, 37(2):95--99, 1997.

\bibitem{Pasquier1996}
Florence Pasquier, Didier Leys, Jan~G.E. Weerts, Francois Mounier-Vehier,
  Frederik Barkhof, and Philip Scheltens.
\newblock {Inter-and intraobserver reproducibility of cerebral atrophy
  assessment on mri scans with hemispheric infarcts}.
\newblock {\em European Neurology}, 36(5):268--272, 1996.

\bibitem{Bresciani2005}
Lorena Bresciani, Roberta Rossi, Cristina Testa, Cristina Geroldi, Samantha
  Galluzi, Mikko~P. Laakso, Alberto Beltramello, Hilkka Soininen, and
  Giovanni~B. Frisoni.
\newblock {Visual assessment of medial temporal atrophy on MR films in
  Alzheimer's disease: Comparison with volumetry}.
\newblock {\em Aging clinical and experimental research}, 17(May 2014):8--13,
  2005.

\bibitem{Henneman2009}
Wouter~J.P. Henneman, Jasper~D. Sluimer, Charlotte Cordonnier, Merel~M.E. Baak,
  Philip Scheltens, Frederik Barkhof, and Wiesje~M. {Van Der Flier}.
\newblock {MRI biomarkers of vascular damage and atrophy predicting mortality
  in a memory clinic population}.
\newblock {\em Stroke}, 40(2):492--498, 2009.

\bibitem{Moller2014}
Christiane M{\"{o}}ller, Wiesje~M. {Van Der Flier}, Adriaan Versteeg, Marije~R.
  Benedictus, Mike~P. Wattjes, Esther L G~M Koedam, Philip Scheltens, Frederik
  Barkhof, and Hugo Vrenken.
\newblock {Quantitative regional validation of the visual rating scale for
  posterior cortical atrophy}.
\newblock {\em European Radiology}, 24(2):397--404, 2014.

\bibitem{Ferreira2016}
Daniel Ferreira, Lena Cavallin, Tobias Granberg, Olof Lindberg, Carlos Aguilar,
  Patrizia Mecocci, Bruno Vellas, Magda Tsolaki, Iwona K{\l}oszewska, Hilkka
  Soininen, Simon Lovestone, Andrew Simmons, Lars~Olof Wahlund, and Eric
  Westman.
\newblock {Quantitative validation of a visual rating scale for frontal
  atrophy: associations with clinical status, APOE e4, CSF biomarkers and
  cognition}.
\newblock {\em European Radiology}, 26(8):2597--2610, 2016.

\bibitem{Ferreira2015}
Daniel Ferreira, Lena Cavallin, Elna-Marie Larsson, J-Sebastian. Muehlboeck,
  Patrizia Mecocci, Bruno Vellas, Magda Tsolaki, Iwona K{\l}oszewska, Hilkka
  Soininen, Simon Lovestone, Andrew Simmons, Lars-Olof Wahlund, and Eric
  Westman.
\newblock {Practical cut-offs for visual rating scales of medial temporal,
  frontal and posterior atrophy in Alzheimer's disease and mild cognitive
  impairment}.
\newblock {\em Journal of Internal Medicine}, 278(3):277--290, 2015.

\bibitem{Westman2011b}
Eric Westman, Lena Cavallin, J.~Sebastian Muehlboeck, Yi~Zhang, Patrizia
  Mecocci, Bruno Vellas, Magda Tsolaki, Iwona K{\l}oszewska, Hilkka Soininen,
  Christian Spenger, Simon Lovestone, Andrew Simmons, and Lars~Olof Wahlund.
\newblock {Sensitivity and specificity of medial temporal lobe visual ratings
  and multivariate regional MRI classification in Alzheimer's disease}.
\newblock {\em PLoS ONE}, 6(7), 2011.

\bibitem{Ferreira2017}
Daniel Ferreira, Chlo{\"{e}} Verhagen, Juan~Andr{\'{e}}s
  Hern{\'{a}}ndez-Cabrera, Lena Cavallin, Chun~Jie Guo, Urban Ekman,
  J.~Sebastian Muehlboeck, Andrew Simmons, Jos{\'{e}} Barroso, Lars~Olof
  Wahlund, and Eric Westman.
\newblock {Distinct subtypes of Alzheimer's disease based on patterns of brain
  atrophy: Longitudinal trajectories and clinical applications}.
\newblock {\em Scientific Reports}, 7(April):1--13, 2017.

\bibitem{Elliott2003}
Rebecca Elliott.
\newblock {Executive functions and their disorders}.
\newblock {\em Imaging neuroscience: clinical frontiers for diagnosis and
  management}, 65(March):49--59, 2003.

\bibitem{Zimny2013}
Anna Zimny, Joanna Bladowska, Ma{\l}gorzata Neska, Kamila Petryszyn, Maciej
  Guzi{\'{n}}ski, Pawe{\l} Szewczyk, Jerzy Leszek, and Marek S{\c{a}}siadek.
\newblock {Quantitative MR evaluation of atrophy, as well as perfusion and
  diffusion alterations within hippocampi in patients with Alzheimer's disease
  and mild cognitive impairment.}
\newblock {\em Medical science monitor : international medical journal of
  experimental and clinical research}, 19:86--94, 2013.

\bibitem{Menendez-Gonzalez2014}
Manuel Men{\'{e}}ndez-Gonz{\'{a}}lez, Alfonso L{\'{o}}pez-Mu{\~{n}}iz,
  Jos{\'{e}}~A. Vega, Jos{\'{e}}~M. Salas-Pacheco, and Oscar
  Arias-Carri{\'{o}}n.
\newblock {MTA index: A simple 2D-method for assessing atrophy of the medial
  temporal lobe using clinically available neuroimaging}.
\newblock {\em Frontiers in Aging Neuroscience}, 6(MAR):1--6, 2014.

\bibitem{Lotjonen2017}
Jyrki Lotjonen, Juha Koikkalainen, Hanneke F M~Rhodius Meester, Wiesje M
  Van~Der Flier, Philip Scheltens, Frederik Barkhof, and Timo Erkinjuntti.
\newblock {Computed rating scales for cognitive disorders from MRI}.
\newblock {\em Alzheimer's {\&} Dementia: The Journal of the Alzheimer's
  Association}, 13(7):P1108, 2017.

\bibitem{LeCun2015}
Yann Lecun, Yoshua Bengio, and Geoffrey Hinton.
\newblock {Deep learning}.
\newblock {\em Nature}, 521(7553):436--444, 2015.

\bibitem{Shen2017a}
Dinggang Shen, Guorong Wu, Heung-il Suk, and Cognitive Engineering.
\newblock {Deep Learning in Medical Image Analysis}.
\newblock {\em Annual Review of Biomedical Engineering}, 19(1):221--248, 2017.

\bibitem{Litjens2017}
Geert Litjens, Thijs Kooi, Babak~Ehteshami Bejnordi, Arnaud Arindra~Adiyoso
  Setio, Francesco Ciompi, Mohsen Ghafoorian, Jeroen~A.W.M. van~der Laak, Bram
  van Ginneken, and Clara~I. S{\'{a}}nchez.
\newblock {A survey on deep learning in medical image analysis}.
\newblock {\em Medical Image Analysis}, 42(December 2012):60--88, 2017.

\bibitem{Esteva2017}
Andre Esteva, Brett Kuprel, Roberto~A. Novoa, Justin Ko, Susan~M. Swetter,
  Helen~M. Blau, and Sebastian Thrun.
\newblock {Dermatologist-level classification of skin cancer with deep neural
  networks}.
\newblock {\em Nature}, 542(7639):115--118, 2017.

\bibitem{Kooi2017}
Thijs Kooi, Geert Litjens, Bram van Ginneken, Albert Gubern-M{\'{e}}rida,
  Clara~I. S{\'{a}}nchez, Ritse Mann, Ard den Heeten, and Nico Karssemeijer.
\newblock {Large scale deep learning for computer aided detection of
  mammographic lesions}.
\newblock {\em Medical Image Analysis}, 35:303--312, 2017.

\bibitem{Gulshan2016}
Varun Gulshan, Lily Peng, Marc Coram, Martin~C. Stumpe, Derek Wu, Arunachalam
  Narayanaswamy, Subhashini Venugopalan, Kasumi Widner, Tom Madams, Jorge
  Cuadros, Ramasamy Kim, Rajiv Raman, Philip~C. Nelson, Jessica~L. Mega, and
  Dale~R. Webster.
\newblock {Development and validation of a deep learning algorithm for
  detection of diabetic retinopathy in retinal fundus photographs}.
\newblock {\em JAMA - Journal of the American Medical Association},
  316(22):2402--2410, 2016.

\bibitem{Roy2017}
Snehashis Roy, John~A. Butman, and Dzung~L. Pham.
\newblock {Robust skull stripping using multiple MR image contrasts insensitive
  to pathology}.
\newblock {\em NeuroImage}, 146(November 2016):132--147, 2017.

\bibitem{Kleesiek2016}
Jens Kleesiek, Gregor Urban, Alexander Hubert, Daniel Schwarz, Klaus
  Maier-Hein, Martin Bendszus, and Armin Biller.
\newblock {Deep MRI brain extraction: A 3D convolutional neural network for
  skull stripping}.
\newblock {\em NeuroImage}, 129:460--469, 2016.

\bibitem{Cole2017}
James~H. Cole, Rudra~P.K. Poudel, Dimosthenis Tsagkrasoulis, Matthan~W.A. Caan,
  Claire Steves, Tim~D. Spector, and Giovanni Montana.
\newblock {Predicting brain age with deep learning from raw imaging data
  results in a reliable and heritable biomarker}.
\newblock {\em NeuroImage}, 163(July):115--124, 2017.

\bibitem{Chen2017}
Hao Chen, Qi~Dou, Lequan Yu, Jing Qin, and Pheng~Ann Heng.
\newblock {VoxResNet: Deep voxelwise residual networks for brain segmentation
  from 3D MR images}.
\newblock {\em NeuroImage}, pages 1--9, 2017.

\bibitem{Wang2018}
Yan Wang, Biting Yu, Lei Wang, Chen Zu, David~S. Lalush, Weili Lin, Xi~Wu,
  Jiliu Zhou, Dinggang Shen, and Luping Zhou.
\newblock {3D conditional generative adversarial networks for high-quality PET
  image estimation at low dose}.
\newblock {\em NeuroImage}, 2018.

\bibitem{Pinto2016}
Adriano Pinto, Victor Alves, and Carlos~A Silva.
\newblock {Brain Tumor Segmentation using Convolutional Neural Networks in MRI
  Images}.
\newblock {\em IEEE Transactions on Medical Imaging}, 35(5):1240--1251, 2016.

\bibitem{Zhao2016}
Liya Zhao and Kebin Jia.
\newblock {Multiscale CNNs for Brain Tumor Segmentation and Diagnosis}.
\newblock {\em Computational and Mathematical Methods in Medicine}, 2016, 2016.

\bibitem{Hosseini-Asl2016}
Ehsan Hosseini-Asl, Robert Keynton, and Ayman El-Baz.
\newblock {Alzheimer's disease diagnostics by adaptation of 3D convolutional
  network}.
\newblock {\em Proceedings - International Conference on Image Processing,
  ICIP}, 2016-Augus(502):126--130, 2016.

\bibitem{Payan2015}
Adrien Payan and Giovanni Montana.
\newblock {Predicting Alzheimer's disease: a neuroimaging study with 3D
  convolutional neural networks}.
\newblock {\em arXiv preprint}, pages 1--9, 2015.

\bibitem{Suk2016}
Heung~Il Suk, Chong~Yaw Wee, Seong~Whan Lee, and Dinggang Shen.
\newblock {State-space model with deep learning for functional dynamics
  estimation in resting-state fMRI}.
\newblock {\em NeuroImage}, 129:292--307, 2016.

\bibitem{Liu2018}
Manhua Liu, Danni Cheng, Kundong Wang, Yaping Wang, and the Alzheimer's
  Disease~Neuroimaging Initiative.
\newblock {Multi-Modality Cascaded Convolutional Neural Networks for
  Alzheimer's Disease Diagnosis}.
\newblock {\em Neuroinformatics}, pages 1--14, 2018.

\bibitem{Donahue2015}
Jeff Donahue, Lisa~Anne Hendricks, Marcus Rohrbach, Subhashini Venugopalan,
  Sergio Guadarrama, Kate Saenko, and Trevor Darrell.
\newblock {Long-term Recurrent Convolutional Networks for Visual Recognition
  and Description}.
\newblock {\em arXiv}, pages 1--14, 2015.

\bibitem{Ypsilantis2016}
Petros-Pavlos Ypsilantis and Giovanni Montana.
\newblock {Recurrent Convolutional Networks for Pulmonary Nodule Detection in
  CT Imaging}.
\newblock {\em arXiv preprint}, pages 1--36, 2016.

\bibitem{Poudel2017}
Rudra~P.K. Poudel, Pablo Lamata, and Giovanni Montana.
\newblock {Recurrent fully convolutional neural networks for multi-slice MRI
  cardiac segmentation}.
\newblock {\em Lecture Notes in Computer Science (including subseries Lecture
  Notes in Artificial Intelligence and Lecture Notes in Bioinformatics)}, 10129
  LNCS:83--94, 2017.

\bibitem{Ferreira2018}
Daniel Ferreira, Sara Shams, Lena Cavallin, Matti Viitanen, Juha Martola,
  Tobias Granberg, Mana Shams, Peter Aspelin, Maria Kristoffersen-Wiberg,
  Agneta Nordberg, Lars~Olof Wahlund, and Eric Westman.
\newblock {The contribution of small vessel disease to subtypes of Alzheimer's
  disease: a study on cerebrospinal fluid and imaging biomarkers}.
\newblock {\em Neurobiology of Aging}, 70:18--29, 2018.

\bibitem{Lindberg2009}
O.~Lindberg, P.~{\"{O}}stberg, B.~B. Zandbelt, J.~{\"{O}}berg, Y.~Zhang,
  C.~Andersen, J.~C.L. Looi, N.~Bogdanovi{\'{c}}, and Lars~Olof Wahlund.
\newblock {Cortical morphometric subclassification of frontotemporal lobar
  degeneration}.
\newblock {\em American Journal of Neuroradiology}, 30(6):1233--1239, 2009.

\bibitem{Muehlboeck2014}
J-Sebastian Muehlboeck, Eric Westman, and Andrew Simmons.
\newblock {TheHiveDB image data management and analysis framework}.
\newblock {\em Frontiers in Neuroinformatics}, 7(January):49, 2014.

\bibitem{Paszke2017}
Adam Paszke, Gregory Chanan, Zeming Lin, Sam Gross, Edward Yang, Luca Antiga,
  and Zachary Devito.
\newblock {Automatic differentiation in PyTorch}.
\newblock {\em 31st Conference on Neural Information Processing Systems},
  (Nips):1--4, 2017.

\bibitem{Jenkinson2002}
Mark Jenkinson, Peter Bannister, Michael Brady, and Stephen Smith.
\newblock {Improved optimization for the robust and accurate linear
  registration and motion correction of brain images}.
\newblock {\em NeuroImage}, 17(2):825--841, 2002.

\bibitem{Jenkinson2001}
Mark Jenkinson and Stephen Smith.
\newblock {A global optimisation method for robust affine registration of brain
  images}.
\newblock {\em Medical Image Analysis}, 5(2):143--156, 2001.

\bibitem{Greve2009}
Douglas~N. Greve and Bruce Fischl.
\newblock {Accurate and robust brain image alignment using boundary-based
  registration}.
\newblock {\em NeuroImage}, 48(1):63--72, 2009.

\bibitem{Wang2017}
Fei Wang, Mengqing Jiang, Chen Qian, Shuo Yang, Cheng Li, Honggang Zhang,
  Xiaogang Wang, Xiaoou Tang, and Sensetime~Group Limited.
\newblock {Residual Attention Network for Image Classification}.
\newblock {\em arXiv preprint}, (1), 2017.

\bibitem{He2015}
Kaiming He, Xiangyu Zhang, Shaoqing Ren, and Jian Sun.
\newblock {Deep Residual Learning for Image Recognition}.
\newblock {\em 2016 IEEE Conference on Computer Vision and Pattern Recognition
  (CVPR)}, (1):770--778, 2016.

\bibitem{Xu2015}
Kelvin Xu, Jimmy Ba, Ryan Kiros, Kyunghyun Cho, Aaron Courville, Ruslan
  Salakhutdinov, Richard Zemel, and Yoshua Bengio.
\newblock {Show, Attend and Tell: Neural Image Caption Generation with Visual
  Attention}.
\newblock In {\em International Conference on Machine Learning}, 2015.

\bibitem{Ba2015}
Jimmy Ba, Volodymyr Mnih, and Koray Kavukcuoglu.
\newblock {Multiple Object Recognition with Visual Attention}.
\newblock {\em arXiv preprint}, pages 1--10, 2015.

\bibitem{Hochreiter1997}
Sepp Hochreiter and J{\"{u}}rgen Schmidhuber.
\newblock {Long short-term memory}.
\newblock {\em Neural Computation}, 9:1735--1780, 1997.

\bibitem{Gers2000}
Felix Gers and Fred Cummins.
\newblock {Learning to Forget: Continual Prediction with LSTM}.
\newblock {\em Neural Computation}, (October), 2000.

\bibitem{Simonyan2015}
Karen Simonyan and Andrew Zisserman.
\newblock {Very Deep Convolutional Networks for Large-Scale Image Recognition}.
\newblock {\em arXiv}, sep 2015.

\bibitem{Loshchilov2016}
Ilya Loshchilov and Frank Hutter.
\newblock {SGDR: Stochastic Gradient Descent with Warm Restarts}.
\newblock {\em arXiv}, pages 1--16, 2016.

\bibitem{Huang2017}
Gao Huang, Yixuan Li, Geoff Pleiss, Zhuang Liu, John~E. Hopcroft, and Kilian~Q.
  Weinberger.
\newblock {Snapshot Ensembles: Train 1, get M for free}.
\newblock {\em arXiv}, pages 1--14, 2017.

\bibitem{Tustison2010}
Nicholas~J Tustison, Brian~B Avants, Philip~A Cook, {Yuanjie Zheng}, Alexander
  Egan, Paul~A Yushkevich, and James~C Gee.
\newblock {N4ITK: Improved N3 Bias Correction}.
\newblock {\em IEEE Transactions on Medical Imaging}, 29(6):1310--1320, 2010.

\bibitem{Buda2017}
Mateusz Buda, Atsuto Maki, and Maciej~A. Mazurowski.
\newblock {A systematic study of the class imbalance problem in convolutional
  neural networks}.
\newblock {\em arXiv preprint}, pages 1--23, 2017.

\bibitem{Cavallin2012a}
Lena Cavallin, Lena Bronge, Yi~Zhang, Anne~Rita {\O}ksengard, Lars~Olof
  Wahlund, Laura Fratiglioni, and Rimma Axelsson.
\newblock {Comparison between visual assessment of MTA and hippocampal volumes
  in an elderly, non-demented population}.
\newblock {\em Acta Radiologica}, 53(5):573--579, 2012.

\bibitem{Velickaite2017}
V.~Velickaite, D.~Ferreira, L.~Cavallin, L.~Lind, H.~Ahlstr{\"{o}}m,
  L.~Kilander, E.~Westman, and E.~M. Larsson.
\newblock {Medial temporal lobe atrophy ratings in a large 75-year-old
  population-based cohort: gender-corrected and education-corrected normative
  data}.
\newblock {\em European Radiology}, pages 1--9, 2017.

\bibitem{Smilkov2017}
Daniel Smilkov, Nikhil Thorat, Been Kim, Fernanda Vi{\'{e}}gas, and Martin
  Wattenberg.
\newblock {SmoothGrad: removing noise by adding noise}.
\newblock {\em arXiv preprint}, 2017.

\bibitem{Landis1977}
J.~Richard Landis and Gary~G. Koch.
\newblock {The Measurement of Observer Agreement for Categorical Data}.
\newblock {\em Biometrics}, 33(1):159, 1977.

\bibitem{Falahati2015}
Farshad Falahati, Seyed~Mohammad Fereshtehnejad, Dorota Religa, Lars~Olof
  Wahlund, Eric Westman, and Maria Eriksdotter.
\newblock {The use of MRI, CT and lumbar puncture in dementia diagnostics: Data
  from the svedem registry}.
\newblock {\em Dementia and Geriatric Cognitive Disorders}, 39:81--91, 2015.

\bibitem{MenendezGonzalez2016}
Manuel {Men{\'{e}}ndez Gonz{\'{a}}lez}, Esther Su{\'{a}}rez-Sanmartin, Ciara
  Garc{\'{i}}a, Pablo Mart{\'{i}}nez-Camblor, Eric Westman, and Andy Simmons.
\newblock {Manual Planimetry of the Medial Temporal Lobe Versus Automated
  Volumetry of the Hippocampus in the Diagnosis of Alzheimer's Disease}.
\newblock {\em Cureus}, 8(3), 2016.

\bibitem{Rhodius-Meester2017}
Hanneke~F.M. Rhodius-Meester, Marije~R. Benedictus, Mike~P. Wattjes, Frederik
  Barkhof, Philip Scheltens, Majon Muller, and Wiesje~M. van~der Flier.
\newblock {MRI visual ratings of brain atrophy and white matter
  hyperintensities across the spectrum of cognitive decline are differently
  affected by age and diagnosis}.
\newblock {\em Frontiers in Aging Neuroscience}, 9(MAY):1--12, 2017.

\end{thebibliography}
\end{document}